\documentclass[12pt,letterpaper]{article}
\usepackage{graphicx}
\usepackage{dcolumn}
\usepackage{bm}
\usepackage{amssymb}
\usepackage{amsmath}
\usepackage{amsfonts}
\usepackage{setspace}
\usepackage[left=1.25in, right=1.25 in]{geometry}
\begin{document}

\title{Veselago Lens for Electrons: \\ Focusing and Caustics in
\\  Graphene \textit{p-n} Junctions. }

\author{ Vadim~V.~Cheianov$^1$,~Vladimir~Fal'ko$^1$, and B.L.
Altshuler$^{2,3}$}
% \date{\today}
\maketitle

\footnotetext[1]{Physics Department, Lancaster University, Lancaster
LA1 4YB, UK } \footnotetext[2]{Physics Department, Columbia
University, 538 West 120th Street, New York, NY 10027, USA}
\footnotetext[3]{NEC-Laboratories America, 4 Independence Way,
Princeton, NJ 085540, USA}

\begin{abstract}
The focusing of electric current by a single \textit{p-n} junction
in graphene is predicted. We show that precise focusing can be
achieved by fine-tuning the densities of carriers on the n- and
p-sides of the junction to equal values, whereas the current
distribution in junctions with different densities resembles
caustics in optics. This finding can be utilized in the engineering
of electronic lenses and focused beam-splitters using
gate-controlled \textit{n-p-n} junctions in graphene-based
transistors.
\end{abstract}
\newpage
 A lot of similarity exists between optics and electronics. Rays
in geometrical optics are analogous to classical trajectories of
electrons, while electron de Broglie waves interfere akin light. The
electron microscope is one example of the technological
implementation of this similarity. The analogy with optics may also
hold a significant potential for semiconductor electronics. In
optics, transparent interfaces between materials are used in lenses
and prisms to manipulate light beams. So far, interfaces have played
a rather different role in semiconductor electronics, where the
central place was, for a long time, occupied by the \textit{p-n
}junction (PNJ). Due to a depletion region near the contact between
two semiconductors with different types of charge carriers (and a
large energy gap), conventional PNJs are not suitable for precision
manipulation of electron beams, which, if realized, may lead to a
new functionality in microelectronics. From this perspective, a lot
of promise is offered by a recently discovered \cite {Novoselov}
truly two-dimensional gapless semiconductor - graphene \cite
{Saito}. Fine-tuning of the carrier density\ in graphene by means of
gates \cite{Geim,Zhang,QHEbi} or doping of the underlying substrate
\cite{Eli} was demonstrated, thus, paving the way towards
controllable ballistic PNJs. On the one hand, the PNJ in graphene is
highly transparent for the charge carriers \cite%
{CheianovFalko,GeimKatsnelson}. On the other, as we show below, the
transmission of electrons through the \textit{p-n} interface
resembles optical refraction \cite{Veselago} at the surface of
metamaterials with negative refractive index
\cite{Pendry,Pendry1,Pendry2}: the straight interface is able to
focus electric current whereas a ballistic stripe of \textit{p-}type
graphene separating two \textit{n-}type regions acts as a lens.

The unique feature of the band structure of graphene (monolayer of
graphite \cite{Saito,DD}) is that its valence band ($\pi $) and
conduction band\ ($ \pi ^{\ast }$) touch each other. In the absence
of doping the Fermi level in graphene is at the energy which belongs
to the both bands and corresponds to the Bloch states in the corners
of the hexagonal Brillouin zone of this two-dimensional honeycomb
crystal. For the states with a small quasimomentum $\hbar \vec{k}$
counted from the corresponding corner of the Brillouin zone, the
dispersion $ \varepsilon (\vec{k})$ and group velocity $\vec{V}=$
$d\varepsilon /d(\hbar \vec{k})$ of electrons are given by
\begin{eqnarray*}
\varepsilon _{c}(\vec{k}) &=&\hbar
vk,\;\vec{V_c}=v\vec{k}/k,\;\text{in
conduction band,} \\
\varepsilon _{v}(\vec{k}) &=&-\hbar
vk,\;\vec{V_v}=-v\vec{k}/k,\;\text{in valence band.}
\end{eqnarray*}%
Figure~1 illustrates  such a dispersion for electrons in
\textit{n-}type (on the left) and \textit{p-} type graphene (on the
right). In a split-gate structure sketched in Fig.~1, voltages $\pm
U$ applied to the two gates shift the degeneracy point of the
electron dispersion cones down, by $\hbar vk_{c}$ on the left and
up, by $\hbar vk_{v}$ on the right and, thus, form a PNJ separating
the \textit{n-} region with the density of electrons
$\rho_{e}=k_{c}^{2}/\pi $  and the \textit{p-}region with the
density of holes  $\rho_{h}=k_{v}^{2}/\pi $. Here $k_{c(v)}$ is the
radius of the Fermi circle in the conduction (valence) band.

The transmission of charge by the PNJ bears striking resemblance to
the refraction of light by left-handed metamaterials
\cite{Pendry,Pendry1,Pendry2} with refractive index equal to $-1$.
As a wave enters such a material, the relative direction of its
group velocity $\vec{V}$ and the wave vector $\vec{k}$ of the wave
reverses, from parallel (in vacuum) to anti-parallel. Therefore,
upon refraction, the sign of the tangential velocity component of
the propagating wave inverts, while the normal component remains the
same. As a result, rays which diverge in vacuum become convergent
after entering the metamaterial \cite{Veselago}. A similar incident
occurs with electrons in the PNJ where the Fermi momentum of the
charge carriers plays the same role as the refractive index in
geometrical optics, with the sign determined by the type of band:
positive for the conduction band and negative for the valence band.

Indeed, let us consider [with reference to Figs.~2(A) and 3(A)]
a de Broglie wave of an electron approaching the PNJ from the \textit{%
n-}side with velocity $\vec{V}_{c}\mathbf{=}(v\cos \theta _{c},v\sin \theta
_{c})$ and $\vec{k}_{c}=(k_{c}\cos \theta _{c},k_{c}\sin \theta _{c})$. At
the interface, this wave is partly reflected to the state with $\vec{k}%
_{c}^{\prime }=(-k_{c}\cos \theta _{c},k_{c}\sin \theta _{c})$ and
partly transmitted to the valence band state with
$\vec{V}_{v}\mathbf{=}(v\cos \theta _{v},v\sin \theta _{v})$ and
$\vec{k}_{v}=(-k_{v}\cos \theta _{v},-k_{v}\sin \theta _{v})$ on the
\textit{p-}side. The probability of the transmission is $\cos
^{2}\theta _{c}/\cos ^{2}(\frac{1}{2}\theta _{c}+\frac{ 1}{2}\theta
_{v})$ \cite{CheianovFalko,GeimKatsnelson}. The component of the
electron momentum along a straight interface should be conserved.
Accordingly, $k_{c}\sin \theta _{c}=-k_{v}\sin \theta _{v}$, that
is, the transmission of electrons is governed by Snell's law:
\begin{equation}
\frac{\sin \theta _{c}}{\sin \theta _{v}}=-\frac{k_{v}}{k_{c}}\equiv n.
\label{refraction-index}
\end{equation}
The negative sign of $n$ in Eq. (\ref{refraction-index}) implies
that the \textit{n-p} interface transforms a divergent flow of
electrons emitted by a source on the \textit{n-}side into a
convergent flow on the \textit{p-}side. This results in focusing
illustrated in Fig.~2(A) for a symmetric junction, $
\rho_{h}=\rho_{e}$ corresponding to $n=-1$. Under the latter
condition electrons injected at $(-a,0)$ in the \textit{n-}region at
the Fermi energy meet again in a symmetric spot at $(a,0)$.

In an asymmetric junction such as shown in Fig.~3(A) for $n=-0.82$
(which corresponds to $\rho_{h}/\rho_{e}=0.67$)\ a sharp focus
transforms into a pair of caustics which coalesce in a cusp - a
singularity in the density of classical trajectories. Similar
singularities in the density of rays, as well as the interference
patterns formed in their vicinity were investigated in optics and
classified \cite{Berry} using the general catastrophe theory
\cite{Arnold}. Ballistic trajectories of electrons in the
\textit{p-}region of an asymmetric PNJ are rays $y=a\tan \theta
_{c}+x\tan \theta _{v}$, where $\theta _{v}$ is related to $\theta
_{c}$ by Eq. (\ref{refraction-index}). The condition for a
singularity, $
\partial y/\partial \theta _{c}=0$ determines the form of caustics $y_{\mathrm{caust}}(x)$ as
well as the position $x_{\mathrm{cusp}}$ of the cusp,
\begin{equation}
y_{\mathrm{caust}} (x)=\pm \sqrt{\frac{\lbrack
x^{2/3}-x_{\mathrm{cusp}}^{2/3}]^{3}}{n^{2}-1}}
,\;x_{\mathrm{cusp}}=|n|a.  \label{cusp}
\end{equation}

To detect focusing by a single flat interface in graphene one can
use a small electric contact as a source of electrons, while another
local probe located on the $p$-side can play the role of a detector.
Electric conductance between the two contacts would reflect the
probability for a carrier to get from the source to the probe. When
the concentration of carriers is low their de Broglie wavelength is
big enough for it to be not impossible to make contacts smaller than
the wavelength. To study  electron transmission in a phase-coherent
system between contacts of such a small size the above-described
classical picture should be complemented with the analysis of
quantum interference pattern of electron de Broglie waves.
Figs.~2(B) and 3(B) visualize the result of full quantum mechanical
calculations of the current of electrons emitted at $(-a,0)$ and
detected by a point contact: near the focal point in the symmetric
PNJ [Fig.~2(B)] and in the vicinity of a cusp [Eq. (\ref{cusp})]
which appears when the symmetry $\rho_{h}=\rho_{e}$ is lifted off
[Fig.~3(B)].  The calculation was performed by applying the Kubo
formula to the single-particle Dirac-like Hamiltonian
\cite{CheianovFalkoFO} of electrons in graphene. Around, but not too
close to the focus ($k_{v}r\gg 1$) the analytically calculated
current is $j\sim (x-a)^{2}/r^{3}$ [$r=\sqrt{(x-a)^{2}+y^{2}}$
stands for the distance from the probe to the focus]. The anisotropy
of the current distribution is caused by the dependence of the
transmission coefficient on the incidence angle and is smeared at
shorter distances $k_{v}r<1$. The current map calculated in the
vicinity of the cusp for $\rho_{h}\neq \rho_{e} $ shows
characteristic patterns described
by the canonical diffraction function for this type of wave catastrophe \cite%
{Berry}. The maximum of the current would be when the probe is at
the tip of the cusp, $(|n|a,0)$. The width $y_{\ast }$ of the bright
spot near the cusp [Fig.~3(B)] or the focus [Fig.~2(A)] in the $y$
direction can be estimated
as $y_{\ast }k_{v}\sim \max \left\{ 1,\left( \frac{1}{2}ak_{c}|n^{-1}-n|%
\right) ^{1/4}\right\} $. Note that for a junction with $n>1$ ($\rho_{h}>\rho_{e}$%
) the pattern near the cusp is mirror-reflected as compared to that
shown in Fig.~3(B) for $n<1$.

It has been discovered \cite{Eigler} in the scanning tunneling
microscopy (STM) studies of elliptically shaped corals on the
surface of copper that the presence of an impurity at one focus of
the ellipse is reflected by the STM map in the vicinity of the other
focus. Therefore,  oscillations of the local density of states of
electrons formed around a static local perturbation \cite{Friedel}
can be replicated through focusing by a carefully engineered fence
of atoms. Similarly, focusing of electrons by a PNJ in graphene
could create a 'mirage', which mimics the effect of a perturbation
on the opposite side of the \textit{n-p} interface. Consider, {\it
e.g.}, a small island of a bilayer \cite{McCannFalko,QHEbi}, which
locally distinguishes between two sublattices (A and B) of the
honeycomb lattice for electrons in the surrounding sheet [due to
Bernal stacking of two adjacent monolayers \cite{Bernal,STM}]. It
induces a change in the local electron density of states (LDoS),
which is different on sublattices A and B. The long-range
oscillations of the alternating LDoS can be detected using STM: as a
difference $\delta j_{\mathrm{A-B}}\sim j_{ \mathrm{A-B}}^{(0)}\sin
(2k_{c(v)}r)/r$\ between the tunneling current from the STM tip to
the A and to B sites. Fig.~2(C)\ shows the results (obtained using
the Green functions technique) of a quantum-mechanical analysis of
oscillations of $\delta j_{\mathrm{A-B}}$ around the mirage image of
a bilayer island formed on the other side of symmetric PNJ in the
monolayer sheet. Fig.~2(D) shows the calculated mirage image of a
spike of electrostatic potential (smooth at the scale of the lattice
constant in graphene), which induces LDoS oscillations equal on the
two sublattices. The difference in the sharpness of these two images
is caused by the lack of backscattering off A-B symmetric scatterers
specific to graphene \cite{CheianovFalkoFO}.

Unlike the ideal left-handed metamaterial \cite{Pendry}, focusing in
the PNJ is not perfect. In symmetric junctions it occurs only for
electrons exactly at the Fermi level, and it is spread into caustics
for electrons excited to higher energies.  This also implies that
the results in Figs.~2 and 3 are only valid for low enough
temperature, $T<\hbar v/a$. For electrons with different energies
the patterns in Fig.~2 smear into patterns characteristic for a cusp
in an asymmetric PNJ,\textit{\ e.g.}, with $ \rho_{h}<\rho_{e}$
shown for the same type of perturbations in Figs.~3(C) and (D). In
this respect, a certain reciprocity also exists: electrons with
energy $ \delta \varepsilon =\hbar v(k_{v}-k_{c})$ counted from the
Fermi level would be focused in the asymmetric PNJ.

Focusing of electrons by a sharp \textit{p-n} junction in graphene
can be used to turn the \textit{n-p-n} junction into a Veselago lens
for electrons. In such a device, Fig.~4(A) the density of charge
carriers in the \textit{p-}region (with width $w$) can be controlled
by the top gate. If the densities in the \textit{n-} and
\textit{p-}regions are equal $\rho_{h}=\rho_{e},$ charge carriers
injected into graphene from the contact $S$ \ shown in Fig.~4(A)
would meet again in the focus at the distance $2w$ from the source
[contact $D_{3}$ in Fig.~4(A)]. Varying the gate voltage over the
\textit{p-}region changes the ratio $n^{2}=\rho_{h}/\rho_{e}.$ This
enables one to transform the focus into a cusp displaced by about
$2(|n|-1)w$ along the \textit{x}-axis and, thus, to shift the strong
coupling from the pair of leads SD$_{3}$ to either SD$_{1}$ (for $
\rho_{h}<\rho_{e}$) or SD$_{5}$ (for $\rho_{h}>\rho_{e}$).
Figs.~4(B,C) illustrate another graphene-based device in which a
prism-shaped top-gate may be used as a focusing beam-splitter. For
example, electrons emitted from the contact $B,$ Fig.~4(B),  are
distributed between the contacts $b$ and $\beta $, whereas the
signal sent from the contact $A,$ Fig.~4(C), is replicated into the
pair of contacts $a$ and $\alpha .$ To mention, graphene has
recently been contacted with a superconducting metal and the
Josephson proximity effect through graphene has been observed \cite
{Morpurgo}. Consequently, a beam splitter Fig.~4(B,C) can be used to
experiment with Einstein-Podolsky-Rosen \cite{EPR}  pairs of
particles.

\pagebreak

\clearpage

\noindent {\bf Fig. 1.} Graphene $p$-$n$ junction (PNJ): monolayer
of graphite is placed over the split gate which is used to create
$n$- (left) and $p$-doped (right) regions. The energy diagram shows
the position of the Fermi level with respect to the touching point
of the valence and the conduction bands.

\vskip 12 pt

\noindent {\bf Fig. 2.} Focusing of electrons by symmetric PNJ,
$\rho_h=\rho_e$ (A) Classical trajectories of electrons diverging
from a source at distance $a$ from the junction become convergent
after refraction. (B) Interference-induced pattern in the charge
current near the focal image of the source-contact. (C,D) "Quantum
mirage" in graphene: local density of states oscillations around the
image of a perturbation applied on the other side of PNJ: (C) a
small island of bilayer and (D) potential of a remote Coulomb
charge.

\vskip 12 pt

\noindent {\bf Fig. 3.} Wave singularities in an asymmetric PNJ,
$\rho_h/\rho_e=0.67$ (A) formation of caustics by refracted waves.
(B) Characteristic interference pattern for the current near the
cusp. (C,D) Local density of states oscillations (in the region
between caustics) created by (C) a small island of bilayer and (D) a
remote Coulomb charge on the other side of PNJ.

\vskip 12 pt

\noindent {\bf Fig. 4.} (A) Electron Veselago lens and  (B, C)
prism-shaped focusing beam splitter in the \textit{n-p-n} junction
in graphene-based transistor.

\end{document}